\documentclass[aps,twocolumn,superscriptaddress,showpacs]{revtex4}



\usepackage{natbib}

\usepackage{graphicx}
\usepackage{dcolumn}
\begin{document}


\title{Evolution of correlation structure of industrial indices of US equity markets}

\author{Giuseppe Buccheri}
\affiliation{Scuola Superiore di Catania, 95123 Catania, Italy}

\author{Stefano Marmi}
\affiliation{Scuola Normale Superiore di Pisa, Piazza dei Cavalieri 7, 56126 Pisa, Italy}
\affiliation{C.N.R.S. UMI 3483 - Laboratorio Fibonacci, Piazza dei Cavalieri 7, 56126 Pisa, Italy}

\author{Rosario N. Mantegna}
\affiliation{Center for Network Science and Department of Economics, Central European University, Nador 9, H-1051, Budapest, Hungary}
\affiliation{Dipartimento di Fisica e Chimica, Universit\`a di Palermo, Viale delle Scienze, Ed. 18, I-90128, Palermo, Italy}

\date{\today}

\begin{abstract}
We investigate the dynamics of correlations present between pairs of industry  indices of US stocks traded in US markets by studying correlation based networks and spectral properties of the correlation matrix. The study is performed by using 49 industry index time series computed by K. French and E. Fama during the time period from July 1969 to December 2011 that is spanning more than 40 years. We show that the correlation between industry indices presents both a fast and a slow dynamics. The slow dynamics has a time scale longer than five years showing that a different degree of diversification of the investment is possible in different periods of time. On top to this slow dynamics, we also detect a fast dynamics associated with exogenous or endogenous events. The fast time scale we use is a monthly time scale and the evaluation time period is a 3 month time period. By investigating the correlation dynamics monthly, we are able to detect two examples of fast variations in the first and second eigenvalue of the correlation matrix. The first occurs during the dot-com bubble (from March 1999 to April 2001) and the second occurs during the period of highest impact of the subprime crisis (from August 2008 to August 2009).
\end{abstract}



\maketitle


\section{Introduction}

The  correlation structure of financial asset returns is a key information for several financial activities ranging from portfolio optimization to risk management and derivative pricing. The correlation structure of financial asset returns has been investigated for stock return time series \cite{Laloux-Cizean-Bouchaud-Potters-1999-PRL,Plerou-Gopikrishnan-Rosenow-Amaral-Stanley-1999-PRL,Mantegna-1999-EPJB} (for recent reviews see \cite{Tumminello-Lillo-Mantegna-2010-JEBO,Bouchaud2011}), market index returns of stock exchanges located worldwide \cite{Bonanno-Vandewalle-Mantegna-2000-PRE,Maslov-2001-PA,Drozdz-2001-PA,Coelh-Gilmore-Lucey-Richmond-Hutzler-2007-PA,Gilmore-Lucey-Boscia-2008-PA,Eryigit-Eryigit-2009-PA,Song2011,Sandoval2012} and currency exchange rates \cite{McDonald-Suleman-Williams-Howison-Johnson-2005-PRE}. The correlation of financial assets is investigated both by considering the spectral density of the eigenvalues of the matrix with tools of multivariate analysis and/or random matrix theory \cite{Laloux-Cizean-Bouchaud-Potters-1999-PRL,Plerou-Gopikrishnan-Rosenow-Amaral-Stanley-1999-PRL,Bouchaud2011} and by using the concept of similarity based graph, i.e. the association  of a network to a similarity matrix \cite{Mantegna-1999-EPJB,Onnela-Chakraborti-Kaski-Kertesz-2002-EPJB,Onnela-Chakraborti-Kaski-Kertesz-2003-PA,Tumminello2005PNAS,Tumminello2007,Tumminello-Coronnello-Lillo-Micciche-Mantegna-2007-IJBC,Tumminello-Lillo-Mantegna-2010-JEBO}. In both cases the aim of the analysis is the selection of information present in the correlation matrix. Correlation between pairs of financial assets is observed to fluctuate around a typical value for periods of time sometimes lasting several years or even decades. However, in addition to this long term regularity also a fast dynamics with a timescale of the order of a few months or even less has been detected \cite{Song2011}.

In this paper we investigate the fast (monthly) dynamics of correlation between industry portfolios of US equity markets. These indices, compiled by the two well known  economists Kennet French and Eugene Fama, are widely considered by the economics and finance research communities as reference portfolios for industry portfolios and have been compiled over a very long period of time starting from July 1962.
 
The paper is organized as follows. In Section \ref{S1:Data}, we briefly present the set of investigated data and we discuss the time scales of the dynamics of correlations of industrial indices. In Section \ref{S1:CorrGraph}, we analyze the correlation based graph associated with the correlation matrix computed by using all daily records of the industrial indices. In Section \ref{S1:MutualInfo}, we discuss the monthly dynamics of plenary maximally filtered graphs (PMFGs) and we compare different correlation based networks by using a mutual information measure based on link overlap. In Section \ref{S1:Spectral}, we investigate the dynamics of the largest eigenvalues and eigenvectors associated with monthly correlation matrices. In the last Section we present our conclusions.

\section{Data and time scales}
\label{S1:Data}

In this study, we investigate a set of 49 value weighted industry portfolios of the US equity markets. The complete list of industries is given in the appendix. Data are recorded daily.  The time period investigated is the time period ranging from July 1969 to December 2011 \cite{note1}. We perform our analysis on the daily return $r_i(t)$  where the label $i$ indicates the industry index and $t$ the trading day.  Starting from the return time series we compute the correlation matrix of this multivariate set of data at each month $t_m$ by using past return recorded  during an evaluation time period of $3$ calendar months. For each month $t_m$, we compute the Pearson correlation coefficient
\begin{equation}
  c_{i,j}(t_m)=\frac{\langle[r_i(k)-\mu_i][r_j(k)-\mu_j]\rangle}{\sigma_i \sigma_j},
\end{equation}
where $\mu_i$ and $\mu_j$ are the sample means and $\sigma_i$ and $\sigma_j$ are the standard deviations of the two industry index time series $i$ and $j$ respectively computed during the 3 month evaluation time period. We have chosen a 3 month evaluation time period because this value is the shortest value leaving the correlation matrix positive definite. In fact 3 months are approximately 60 trading days and the number of industry indices is 49. In this way we can investigate the fast dynamics of the correlation matrix by ensuring that all the eigenvalues of the matrix remain positive \cite{Ledoit-Wolf-2004-JMA}.  A similar analysis was performed in the recent investigation of 57 indices of stock markets located all over the world \cite{Song2011}.

\begin{figure}
\includegraphics[scale=0.33]{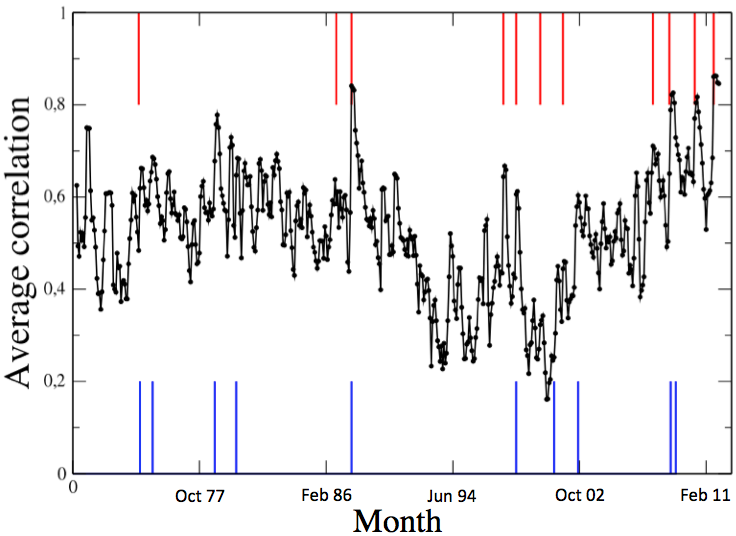}
\caption{Monthly time evolution of the average correlation of the non diagonal elements of correlation matrices estimated by using an evaluation time period of 3 months. The time evolution shows a fast dynamics and an overall slow behavior. Vertical red lines  indicate months when prominent events occurred. Specifically from left to right we have: 1) October 1973 first oil embargo; 2) October 1986 tax reform act; 3) October 1987 stock market crash; 4) October 1997 Asian crisis; 5) August 1998 Russian crisis; 6) March 2000 Nasdaq reaches its maximum value during the dot-com bubble; 7) September 2011 $9/11$ terrorist attack;  8) August 2007 interbank market freezes; 9) September 2008 Lehman's bankruptcy; 9) May 2010 Eurozone and International Monetary Fund decide the first bailout for Greece, and 10) August 2011 onset of the European sovereign debt crisis. Vertical blue lines indicate the months with the top 10 monthly negative returns of the Standard \& Poor's 500.}
\label{fig3m6m}
\end{figure}

In Fig. \ref{fig3m6m} we show the average correlation of the non diagonal elements of monthly correlation matrices estimated for the 49 industry indices. The figure suggests that the fast time scale of the average correlation among indices might be sometime shorter than three trading months. Unfortunately, we cannot use shorter evaluation time periods without altering the positive definite nature of the correlation matrix \cite{Ledoit-Wolf-2004-JMA}. The figure also show that fast changes of the average correlation are detected both in the presence of events that are exogenous to the market  (like, for example, the Asian crisis of 1997 or the sovereign debt crisis of 2011) and in the presence of events endogenous to the market with apparent no external explanation triggering large variation of representative market indices (like, for example,  the market decline of October 1978 or market crash of October 1987). 

\section{Correlation based graphs}
\label{S1:CorrGraph}

\begin{figure}
\includegraphics[scale=0.32]{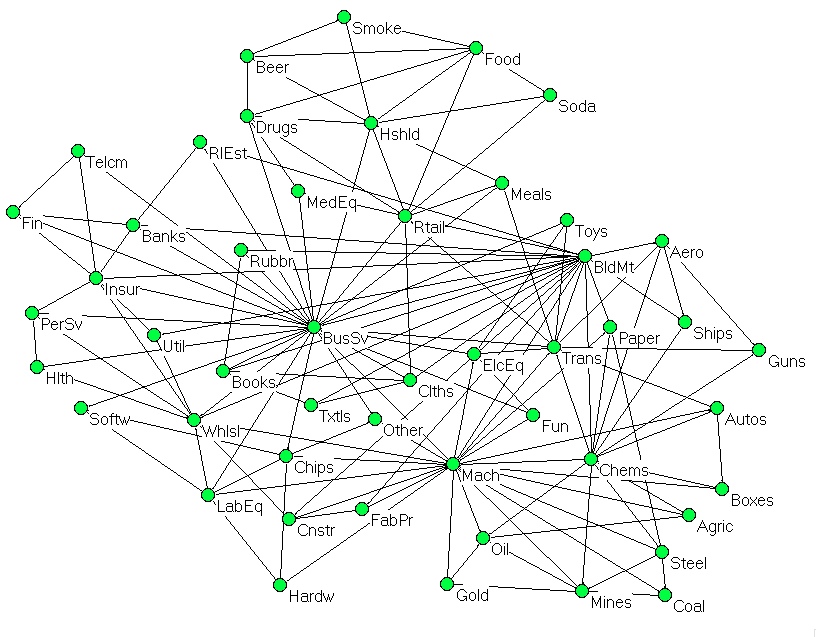}
  \caption{PMFG of the set of 49 industry indices obtained from the correlation matrix of daily index returns estimated by using all daily records during the period from July 1969  to December 2011. The most connected industry indices are BusSv (business services), Mach (machinery) and BldMt (construction materials).}
  \label{figPMFG}
\end{figure}

Correlation based graphs are powerful tools detecting, analyzing and visualizing in an efficient way part of the most statistically robust information which is present in the correlation matrix \cite{Tumminello-Lillo-Mantegna-2010-JEBO}. Here we start our investigation by considering the PMFG \cite{Tumminello2005PNAS} of the 49 industry indices obtained from the correlation matrix estimated by using all the 10621 daily records of the selected time period (1/7/1969 - 31/7/2011). The PMFG obtained for the entire period is shown in Fig. \ref{figPMFG}. It should be noted that industry indices present a tendency to cluster in groups of indices of related economic sectors.

We detect clusters of industry indices present in the PMFG by using the Infomap method proposed by Rosvall and Bergstrom \cite{Rosvall-Bergstrom-2008-PNAS}. This algorithm is considered one of the best algorithms of community detection in networks \cite{Fortunato-2010-PR,Lancichinetti-Fortunato-2009-PRE}. The method uses the probability flow of random walks  to identify the community structure of the system. We repeat the application of the method 100 times to detect a minimum value of the fitness parameter estimating the goodness of the partition and to evaluate the robustness of the partition obtained.

\begin{figure}
\includegraphics[scale=0.3]{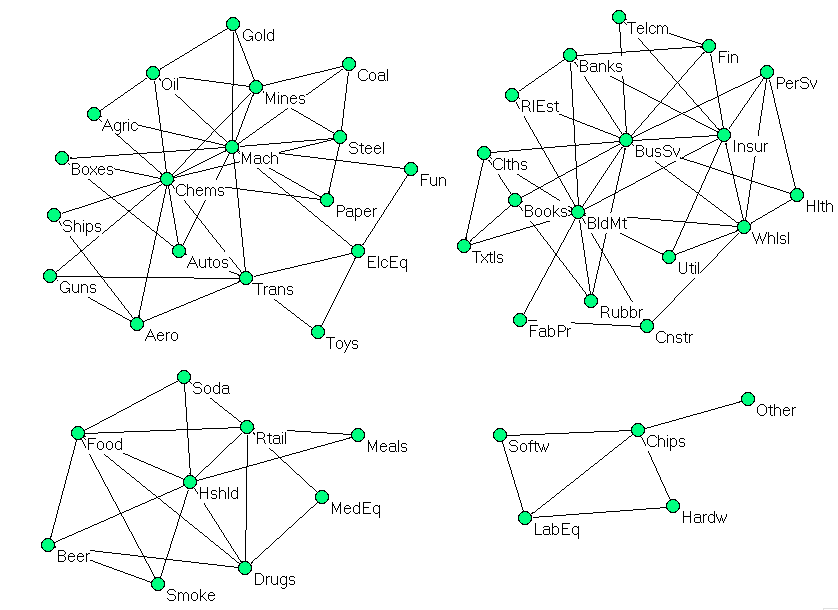}
\caption{Clusters (communities) of industry indices detected into the PMFG (computed by using records for the period from 7/69 to 12/11) by the Infomap community detection algorithm. The algorithm detects four clusters of indices of stocks acting in different sectors of the economy. Top left: $C_2$, in this cluster we have industries dealing with commodities, basic materials and transportation. Top right: $C_1$, industries dealing with financial services, personal and business services, construction and building materials, wholesale, and utilities. Bottom left: $C_3$, food, pharmaceuticals and medical equipments, consumer products and retail.  Bottom right: $C_4$, information technology oriented industrial sectors.}
\label{figInfomap}
\end{figure}

The result of the partitioning of the PMFG is shown in Fig. \ref{figInfomap}. The method identifies four distinct clusters. The top left cluster is a cluster of 18 industry indices dealing with commodities, basic materials and transportation. In the top right one, there are 17 indices of stocks belonging to the sectors of financial services, personal and business services, construction and building materials, wholesale, and utilities. The other two clusters are smaller ones. The one at the bottom left comprises 9 indices of stocks of economic sectors as food, pharmaceuticals and medical equipments, consumer products and retail. The last one at the bottom right is a cluster of 5 indices belonging to the information technology economic sector. In fact it comprises chips manufacturing, hardware and software and laboratory equipment.

\section{Dynamics of the PMFGs}
\label{S1:MutualInfo}

\begin{figure}[t]
\includegraphics[scale=0.35]{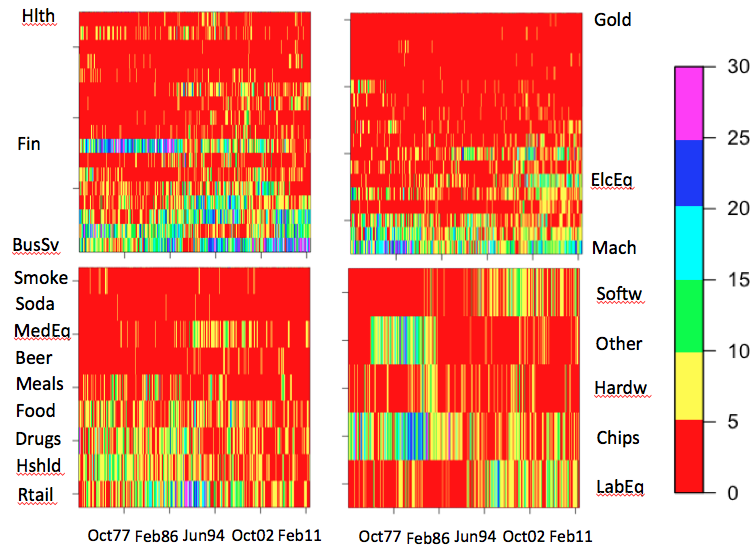}
\caption{Color code representation of the time evolution of the degree of industry indices of the PMFGs computed monthly from Sep 69 to Dec 2011 by using a 3 month evaluation time period. Industry indices are shown in four panels each one representing industry indices of one of the four clusters detected by the Infomap algorithm. Specifically we have grouped industry indices by ordering them from bottom to top in each of the $C_1$ (top left), $C_2$ (top right), $C_3$ (bottom left) and $C_4$ (bottom right) clusters. For clusters $C_3$ and $C_4$ the short names of industry indices are provided in the figure. For clusters $C_1$ and $C_2$, the complete sequence of industry indices is the following from bottom to top: $C_1$: BusSv, BldMt, Whlsl, Insur, Banks, Clths, Books, Fin, Cnstr, PerSv, Rubbr, Telcm, Txtls, RlEst, FabPr, Util, and Hlth; $C_2$: Mach, Chems, Trans, Mines, Steel, ElcEq, Aero, Paper, Autos, Oil, Boxes, Fun, Toys, Coal, Ships, Guns, Agric, and Gold. Within each cluster, industry indices are ordered from bottom to top according to the rank provided by the Infomap algorithm applied to the correlation matrix of the entire period (1969-2011). The color scale is provided on the right of the figure. White spots indicates values higher than 30. The PMFGs subgraphs  of the four clusters are shown in Fig. \ref{figInfomap}.}
\label{figDTE}
\end{figure}

For each month of the time period ranging from September 1969 to December 2011 we obtain a correlation matrix, and from each correlation matrix we construct the associated PMFG.  We therefore investigate how links of the PMFGs change from month to month. Specifically, we consider the time evolution of the degree of each vertex, the time evolution of the vertex betweenness \cite{Newman} and the time evolution of the link mutual information as defined in Ref. \cite{Song2011}. In Fig. \ref{figDTE} we show a color code representation of the monthly time evolution of the degree of each industry index of the PMFGs for the investigated time period (1969-2011). In the figure, different industry indices are ordered according to their rank within the four clusters obtained by the Infomap algorithm during the best partitioning of the PMFG computed by using the correlation matrix estimated using all daily records of the time period 1969-2011 (see Fig. \ref{figPMFG} and Fig. \ref{figInfomap} of Section \ref{S1:CorrGraph}).

\begin{figure}[t]
\includegraphics[scale=0.34]{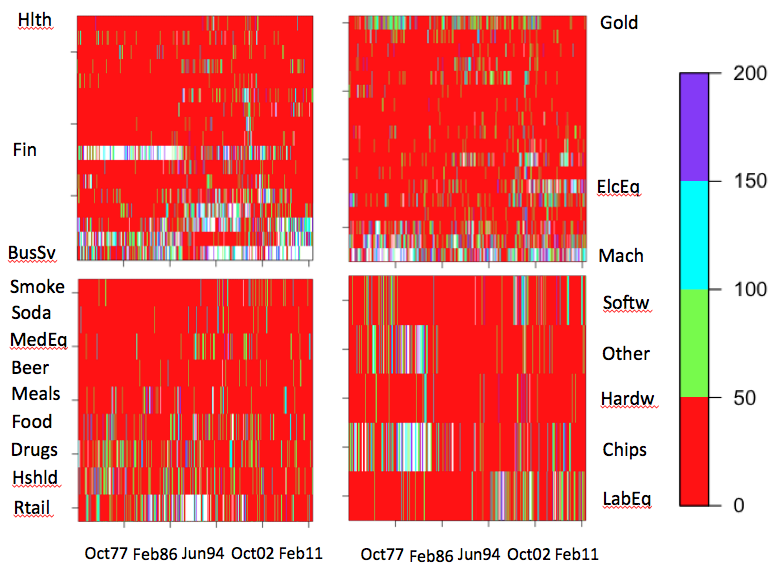}
\caption{Color code representation of the time evolution of the non normalized vertex betweenness of industry indices of the PMFGs computed monthly form Sep 69 to Dec 2011 by using a 3 month evaluation time period. As in the previous figure, industry indices are shown in four panels each one representing industry indices of one of the four clusters detected in the PMFG of the entire time period by the Infomap algorithm. The color scale is provided on the right of the figure. White spots indicates values higher than 200. Industry indices are ordered as in Fig. \ref{figDTE}. The ordering of the indices is the same as in Fig. \ref{figDTE}.}
\label{figBTE}
\end{figure}

Fig. \ref{figDTE} shows that the time evolution of the degree of industry indices is for several indices quite stable over time for a time period as long as 40 years. For example, in cluster $C_2$ (top right panel) the most connected index is the Mach (machinery) index and this index maintains this role over the entire time period. Similarly, one of the least connected indices (Gold) of the cluster also maintains this status over the entire period of time. However, different 
time evolutions of the degree are also observed. For example, in cluster $C_1$ (top left panel) the trading industry index (Fin) is characterized by two different time periods one occurring before 1987 and the other after. Similar time localized periods of high (or low) degree are observed for example (i) for retail industry (Rtail) and medical equipment industry (MedEq) in cluster $C_3$, and (ii) for computer software industry (Softw) and measuring and control equipment industry (LabEq) in cluster $C_4$. These results manifest the fact that  
there is an underlying dynamics of the correlation present between pair of industry indices. The presence of a dynamics observed for the degree is supported and complemented by observing the dynamics of the vertex betweenness, which is another key network indicator of network topology. In fact in Fig. \ref{figBTE} we see, even more clearly than in the case of the degree, that, for some indices, vertex betweenness changes over time. Again a striking example is the behavior of the trading industry index (Fin). For this vertex the betweenness is very high during the period from 1969 to 1987 and decreases significantly after that year. It should be noted that degree and betweenness may carry, in general, different information. This is somewhat evident when one analyze the case of precious metals industry (Gold). For this index (index at the top of cluster $C_2$ in Figs \ref{figDTE} and \ref{figBTE} ) the degree is very low (see Fig. \ref{figDTE}) whereas the vertex betweenness occasionally presents intermediate values (green spots in the time evolution of Fig. \ref{figBTE}) indicating that this index is typically outside the core region of cluster $C_2$ but often acting as a bridge across different clusters of the entire network.

\begin{figure}[b]
 \includegraphics[scale=0.43]{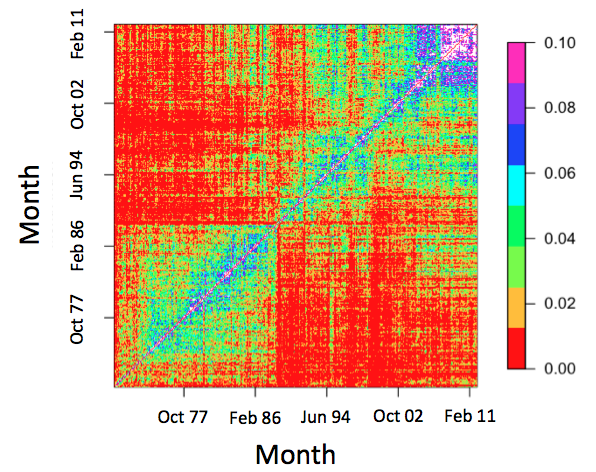}
  \caption{Color code representation of link mutual information between all pairs of a PMFG estimated at month $i$ (raw of the matrix) and a PMFG estimated at month $j$ (column of the matrix). The time increases from left to right and from bottom to top. The color scale is provided on the right of the figure. White spots indicate values of the mutual information higher than 0.1.}
  \label{figMD}
\end{figure}

To obtain a measure of the overlap of a PMFG obtained for a certain calendar month with another one we compute the mutual information of links between all pairs of monthly PMFGs. The mutual information of links was defined in Ref. \cite{Song2011}. The result of this estimation is  shown in the mutual information matrix of Fig. \ref{figMD}. The mutual information matrix presents an approximate block like structure. Specifically, a first block is detected during the period approximately ranging from 1969 to 1987. A second block, more internally structured, is observed from the end of 1987 to 1999 and a third one is observed after 2004. We interpret the first and the last of these periods as periods characterized by a relative high stability of the PMFGs indicating that the relative ordering of the correlation values among the pairs of stocks is approximately maintained. On the other hand, from one period to another and, to a lesser extent, within the second period, transitions from one period to another are observed and imply the variation of the ranking of the most intense correlations detected among pairs of industry indices. For example, Fig. \ref{figDTE} and Fig. \ref{figBTE} show that the trading industry index (Fin) has a quite different behavior before and after 1987. This implies  that its ranking in the ordered list of the correlation similarity measure between pairs of industry indices jumped from high values to relatively low values around the end of 1987.

\subsection{Spearman correlation of the similarity measure}
The results of Fig. \ref{figMD} show that the highest correlation values of pairs of stocks experienced changes in their rank of the Pearson correlation  in 1987, 1999 and in 2004 and perhaps also in other years to a lesser degree. These changes are certainly due to relative variation of the highest values of the Person's correlation because the PMFGs structure is controlled by correlation values that are primarily among the highest values of correlation for each element of the system \cite{Tumminello2007}. It is therefore of interest to evaluate whether there is also a change of the global ranking of the Pearson correlation between pairs of industry indices. 

\begin{figure}
\includegraphics[scale=0.43]{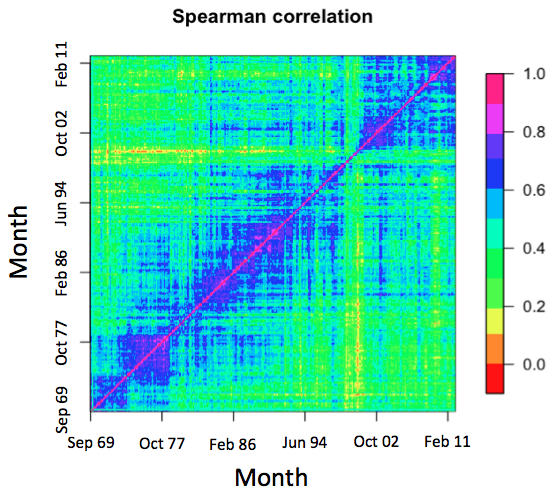}
\caption{Color code representation of Spearman rank correlation between all pairs of rank of Pearson correlation coefficients estimated at month $i$ (raw of the matrix) and rank of Pearson correlation coefficients estimated at month  $j$ (column of the matrix). The time increases from left to right and from bottom to top. The color scale is provided on the right of the figure.
}
\label{figSpearman}
\end{figure}

We consider this problem by measuring the Spearman rank correlation \cite{Spearman1904}, i.e. the Pearson correlation coefficient between the ranked variables associated with the value of the pair correlation of each pair of index indices. The Spearman rank correlation is therefore quantifying the similarity of the ranking of the distinct correlation coefficients measured monthly by the Pearson correlation matrix. In Fig. \ref{figSpearman} we show a color code representation of the Spearman rank correlation observed among all Person's correlation matrices computed for all the investigated months. 
As in the case of link mutual information discussed previously, Fig. \ref{figSpearman} also shows a block like structure but the boundaries of the blocks are, in most cases, observed for times different from the ones characterizing the mutual information (see Fig. \ref{figMD} ). In fact, Fig. \ref{figSpearman} shows major boundaries between different blocks in 1978, 1992, 1999 and 2001. The behavior during the time interval 1999-2001 is markedly different from any other past and future period indicating that during the dot-com bubble the rank of the correlation coefficients among industry indices of the US market was significantly 
different from all other time periods. 
Fig. \ref{figSpearman} also shows that the rank of Pearson correlation coefficient among industry indices of the period 2001-2010 is significantly different from the rank observed in the past, especially when compared with the time period prior to 1978.

\section{Spectral analysis of the correlation matrices}
\label{S1:Spectral}
In the previous section we have seen that both the highest values of the pair correlation and more generally the entire set of Pearson correlation coefficients have changed 
rank over time with abrupt changes in a few cases localized at specific times. Here we continue the analysis of the properties of the Pearson correlation as a function of the time by investigating the spectrum of eigenvalues and eigenvectors of the correlation matrices. In our analysis we mainly focus on the time dynamics of the largest eigenvalues and of their corresponding eigenvectors. In Fig. \ref{figTE} we show the time evolution  of the first, second and third eigenvalue of the monthly correlation matrices. The time profile of the first eigenvalue is highly correlated with time profile of the average correlation (see Fig. {\ref{fig3m6m}) and therefore no additional information can be easily extracted from him. 
The time evolution of the  second eigenvalue is more informative because it shows a few abrupt changes in specific periods of time. In Fig. \ref{figTE} we point out that biggest changes are observed during April-June 1999, March-May 2000, June-March 2001, and July-September 2008. The third eigenvalue has a more limited excursion, its mean value is equal to 1.82 and the standard deviation is 0.46. 
It will therefore be very difficult to extract the information associated with this eigenvalue in a statistically reliable way. In summary, the first two eigenvalues are the only large eigenvalues carrying information that is statistically reliable and easily distinguishable from fluctuations induced in the estimation process by the limited number of records used to compute the correlation matrix \cite{Laloux-Cizean-Bouchaud-Potters-1999-PRL}.

We interpret the spikes observed in the time evolution of the second eigenvalue of the correlation matrix as an indication of changes occurred in the correlation matrix and specifically changes affecting the correlation of some specific sets of industry indices against all the others. 
This conclusion is also consistent with the results obtained investigating the Spearman rank correlation discussed in the previous section.
It should be noted that the main spikes are localized during the time period from April 1999 to March 2001 when the market experienced the strongest manifestation of the dot-com bubble (NASDAQ index reached its maximum value on March 10th, 2000) and its deflation (NASDAQ index declined to half its value within a year from the maximum value). The other most prominent spike is observed for the time period July-September 2008 which was the hottest period of the 2008 financial crisis that led to the Lehman's failure of September 15th,  2008. The spikes of the second eigenvalue are therefore indicating two major crises experienced by the US equity markets in recent years.

\begin{figure}
\includegraphics[scale=0.34]{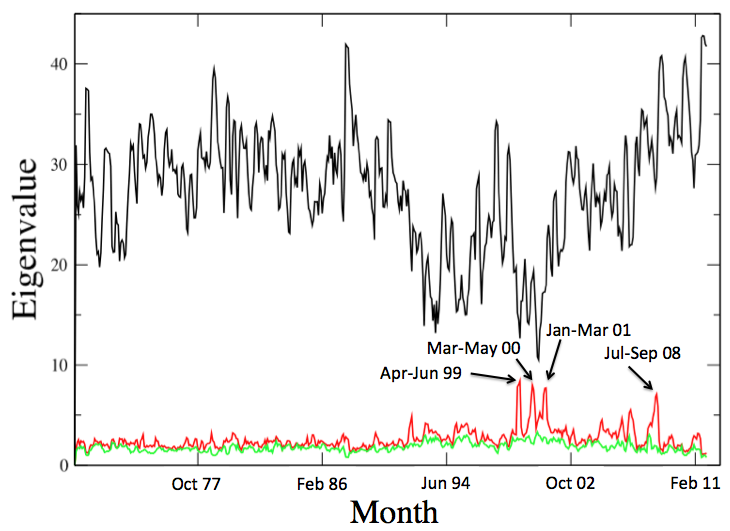}
\caption{Monthly time evolution of the first (black (top) line), second (red (middle) line) and third (green (bottom) line) eigenvalue of correlation matrices computed by using a 3 month evaluation time period. The dates of the most prominent spikes of the second eigenvalue are reported in the figure.}
\label{figTE}
\end{figure}

We investigate the nature of information present in the two eigenvalues by analyzing the profile of the eigenvectors associated with the two largest eigenvalues. In Fig.~\ref{figFE} we show a color code representation of the components of the eigenvector associated with the first eigenvalue for all the 508 investigated months. When the spectral analysis of a matrix is performed, the direction of the eigenvector is arbitrary. In the figure we select the direction associated with a positive component of the eigenvector by setting positive the component of the business services industry index (BusSv). The eigenvector components are almost all positive indicating the presence of a common factor driving all industry indices. The only significant exception is the last industry index of cluster $C_2$. This industry index is the precious metals industry index (Gold). 

The time evolution of the components of the eigenvector of the second eigenvalue has a less straightforward interpretation. During the time period before 1987 clusters $C_1$ and $C_3$ show values of the components of the eigenvector with values preferentially positive or negative respectively (see the top panel of Fig. \ref{figSE}). After 1987 the general behavior is much more complex although some local regularities emerge. For example, one prominent case is observed during the period from March 1999 to April 2001, i.e. the period when the second eigenvalue shows prominent spikes. During this time period the values of the components of the eigenvector of industry indices of the $C_4$ cluster assume, with the only exception of the index {\it Other}, very high positive values while the large majority of all other industry indices assume values close to zero or negative (see the bottom panel of Fig. \ref{figSE}). This is a direct manifestation of the fact that during that period of time there was a strong decoupling of the correlation present between industry indices directly related to information technology, computers (Hardw), computer software (Softw), electronic equipment (Chips), and measuring and control equipment(LabEq) and the rest of indices. 


\begin{figure}
\includegraphics[scale=0.45]{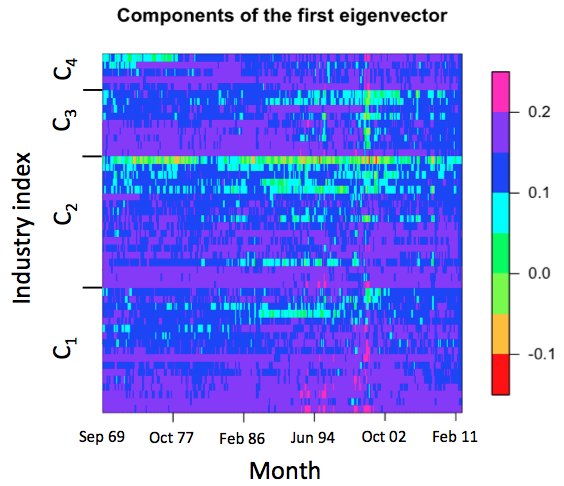}
\caption{Color code representation of the components of the first eigenvector as a function of time (vertical axis). The direction of the eigenvector is selected by making positive the component of the BusSv industry index. Industry indices are ordered from bottom to top according to the clusters detected by the Infomap algorithm in the unconditional PMFGs of Fig.~\ref{figPMFG}. $C_1$ to $C_4$ are the clusters shown in Fig.~\ref{figInfomap}.}
\label{figFE}
\end{figure}

In summary, our analysis of the dynamics of the first two largest eigenvalues shows a strong alteration of the value of eigenvalues for the time period associated with two of the most prominent market periods of the last 40 years, which are the dot-com bubble and the peak of the subprime crisis. Concerning the relevance of the second eigenvalue in terms of explained variance, it is worth noting that, in absolute terms, the second eigenvalue explains a maximal amount of approximately $16 \%$ of the variance during the two periods of time discussed here. Moreover, in relative terms the two periods are quite different because for the dot-com period the first eigenvalue explain approximately $31\%$ of the variance whereas during the 2008-2009 crisis the first eigenvalue explains roughly $70\%$ of the variance. In other words the 2008-2009 crisis affects all the industry indices whereas the dot-com bubble was primarily impacting on the information technology industry sector.

\begin{figure}
\includegraphics[scale=0.36]{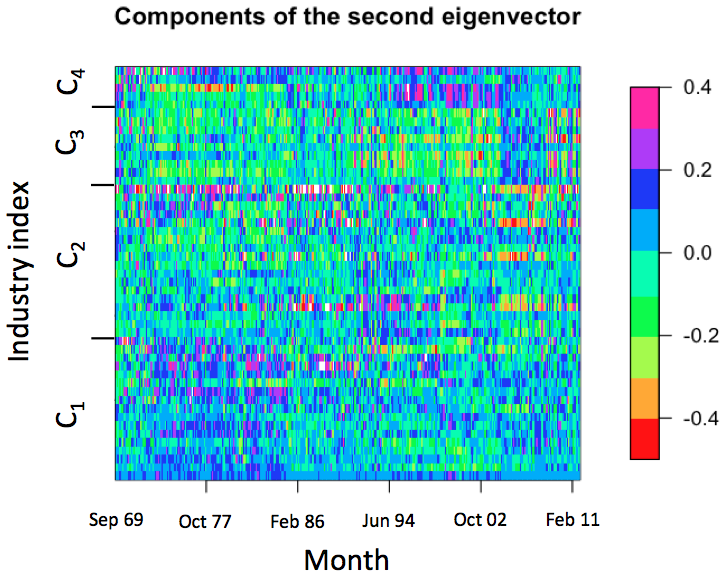}
\includegraphics[scale=0.43]{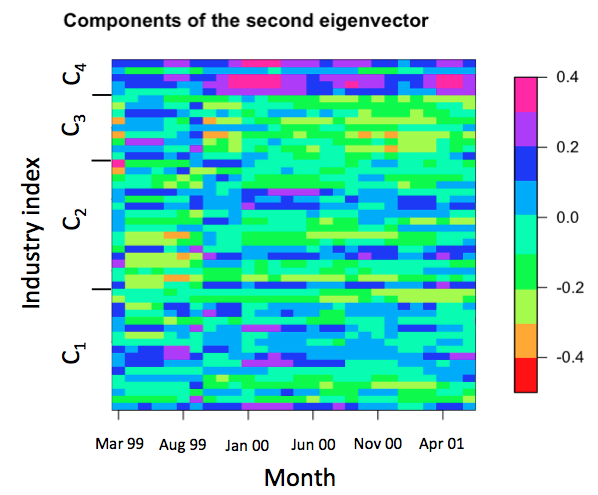}
\caption{Color code representation of the components of the second eigenvector as a function of time. The direction of the eigenvector is selected by making positive the component of the BusSv index. Market indices are ordered from bottom to top according to the clusters detected by the Infomap algorithm in the unconditional PMFGs of Fig. \ref{figPMFG}. $C_1$ to $C_4$ are the clusters shown in Fig. \ref{figInfomap}. The top panel describes the entire time period investigated whereas the bottom panel refers to the dot-com time period from March 1999 to May 2001.}
\label{figSE}
\end{figure}

\section{Conclusions}
\label{S1:Conclusion}

In this paper we investigate the dynamics of correlation present among pairs of industry  indices of US stock traded in US markets.  The study is performed by using 49 industry index time series computed by K. French and E. Fama since 1962.  By investigating this set of industry indices over a period of time spanning more than forty years, we discover that the correlation between industry indices presents both a fast and a slow dynamics. The slow dynamics indicates that a different degree of diversification of the investment is possible in different periods of time and that the time scale of these changes is at least as slow as five years. On top to this slow dynamics, we also detect a fast dynamics associated with exogenous or endogenous market events. Specifically, by computing the correlation matrix each trading month using a 3 month evaluation time period we show that the correlation matrix presents both very long periods of time during which the relative rank of Pearson correlation coefficients between pairs of industry indices is rather stable and periods when there is a significant modification of relative correlation in relatively short periods of time. Two major examples of these abrupt changes are observed during the dot-com bubble (March 1999 to April 2001) and during the period of highest impact of the subprime crisis (August 2008 to August 2009).

\begin{acknowledgments}
The research leading to these results has received partial funding from the European Union, Seventh Framework Programme FP7/2007-2013 under grant agreement CRISIS-ICT-2011-288501.

\end{acknowledgments}

\appendix*\section{Set of market indices}

We investigate the daily synchronous dynamics of 49 industry indices compiled by K. French and E. Fama during the time period June 1969 December 2011. The industry indices investigated are: Agriculture (Agric), Food products (Food), Candy \& Soda  (Soda), Beer \& Liquor (Beer), Tobacco products (Smoke), Recreation (Toys), Entertainment (Fun), Printing and publishing  (Books), Consumer goods (Hshld), Apparel (Clths), Healthcare 	(Hlth), Medical equipment (MedEq), Pharmaceutical products (Drugs), Chemicals (Chems), Rubber and plastic products (Rubbr), Textiles (Txtls), Construction materials (BldMt), Construction (CNstr), Steel works etc. (Steel), Fabricated products (FabPr), Mach (Machinery), Electrical equipment (ElcEq), Automobiles and trucks (Autos), Aircraft (Aero), Shipbuilding, railroad equipment (Ships), Defense (Guns), Precious metals (Gold), Non-metallic and industrial metal mining (Mines), Coal (Coal), Petroleum and natural gas  (Oil), Utilities (Util), Communications (Telcm), Personal services (PerSv), Business services (BusSv), Computers (Hardw), Computer software (Softw), Electronic equipment (Chips), Measuring and control equipment (LabEq), Business supplies (Paper), Shipping containers (Boxes), Transportation (Trans), Wholesale (Whlsl), Retail (Rtail), Restaurants, hotels, motels (Meals), Banking (Banks), Insurance (Insur), Real estate (RlEst), Trading  (Fin), and Others  (Other). The order of the indices, the definition and the descriptive codes are the ones used by French and Fama \cite{Website}.


\begin{thebibliography}{21}
\bibitem{note1}
Indeed the time period of the database starts on July 1963.  However only since July 1969 returns are record for all the 49 indices. During the period from July 1963 to June 1969 many data are missing for several industry indices.
\end{thebibliography}
\end{document}